\documentclass[twocolumn]{aastex631}

\shorttitle{Coronal and chromospheric emission in A-type stars}
\shortauthors{G\"unther et al.}

\graphicspath{{./}{figures/}}

\begin{document}

\title{Coronal and chromospheric emission in A-type stars}

\author[0000-0003-4243-2840]{Hans Moritz G{\"u}nther}
\affiliation{MIT Kavli Institute for Astrophysics and Space Research, 77 Massachusetts Avenue, Cambridge, MA 02139, USA}
\author[0000-0001-9834-7579]{Carl Melis}
\affiliation{Center for Astrophysics and Space Sciences, University of California, San Diego, CA 92093-0424, USA}
\author[0000-0002-4939-8940]{J. Robrade}
\affiliation{Hamburger Sternwarte, Gojenbergsweg 112, 21029 Hamburg, Germany}
\author[0000-0002-5094-2245]{P. C. Schneider}
\affiliation{Hamburger Sternwarte, Gojenbergsweg 112, 21029 Hamburg, Germany}
\author[0000-0002-0826-9261]{Scott J. Wolk}
\affiliation{Smithsonian Astrophysical Observatory, MS 70, 60 Garden St., Cambridge, MA 02138}
\author[0000-0002-9569-2438]{Rakesh K. Yadav}
\affiliation{Department of Earth and Planetary Sciences, Harvard University, Cambridge, MA 02138, USA}



\begin{abstract}
Cool stars on the main sequence generate X-rays from coronal activity, powered by a convective dynamo. With increasing temperature, the convective envelope becomes smaller and X-ray emission fainter. We present Chandra/HRC-I observations of four single stars with early A spectral types. Only the coolest star of this sample, $\tau^3$~Eri ($T_\mathrm{eff}\approx8\,,000$\,K), is detected with $\log(L_X/L_\mathrm{bol})=-7.6$ while the three hotter stars ($T_\mathrm{eff}\geq8\,,000$\,K), namely $\delta$~Leo, $\beta$~Leo, and $\iota$~Cen, remain undetected with upper limits $\log(L_X/L_\mathrm{bol})<-8.4$. The drop in X-ray emission thus occurs in a narrow range of effective temperatures around $\sim 8100$\,K and matches the drop of activity in the \ion{C}{3} and \ion{O}{6} transition region lines.

\end{abstract}



\section{Introduction} \label{sec:intro}
Stars across the main-sequence (MS) produce X-ray emission in two fundamentally different mechanisms. Cool stars have convective envelopes, 
which can produce differential rotation and convective turbulence, giving rise to strong magnetic fields through the dynamo mechanism \citep{2017LRSP...14....4B}. These magnetic fields then help sustain coronal activity around these stars.
Essentially all close-by late-type stars are X-ray emitters \citep{2004A&A...417..651S}. On the other end of the MS the most massive stars have fast winds. Instabilities in the winds heat the gas to a few MK. Again, nearly all of them are X-ray emitters \citep{1996A&AS..118..481B,1997A&A...322..167B}. Stars from mid-A to B operate neither mechanism: Their winds are too weak to produce detectable X-ray emission and their atmospheres are radiatively dominated and do not drive a turbulent convective dynamo. Mid-A to B-type stars thus are X-ray dark \citep{1997A&A...318..215S}.

Nevertheless, some A and B-type stars  are seen in the ROSAT
All-Sky Survey (RASS) or in other X-ray datasets \citep{2020ApJ...902..114W}, because they often have unresolved late-type
companions. Due to the shorter lifetime of the A-type star, the companion is
still at an early stage of its evolution when the A-star is on the MS and thus the companion is X-ray bright. The RASS
catalog contains 312 bright A-type stars. This is a detection rate of 10-15\% \citep{2007A&A...475..677S}. 
In the sub-sample studied by \citet{2000A&A...359..227H}, X-ray hardness and flux are similar to late-type stars indicating unresolved companions are responsible for that emission. \citet{2003A&A...407.1067S} observed five RASS sources with apparent X-rays with \emph{Chandra}. In three targets the X-rays are due to a resolved companion, and \citet{2003A&A...407.1067S} argue based on spectral properties that the remaining targets probably have an unresolved companion that generates the X-ray emission.

However, star spots leading to photometric variability with a low amplitude of 0.05\% have been found for Vega \citep{2015A&A...577A..64B} together with a weak (disk-averaged line-of-sight component $< 1$~G) magnetic field \citep{2009A&A...500L..41L,2010A&A...523A..41P}.
Also, recent studies with \emph{Kepler} and \emph{TESS} do indeed find rotational modulation of early A-type stars \citep{2011MNRAS.415.1691B,2017MNRAS.467.1830B,2019MNRAS.487.4695S} and sometimes signatures of what seem to be magnetic flares \citep{2012MNRAS.423.3420B}. However, the latter can usually be attributed to binarity or artifacts such as contamination of the lightcurve by nearby sources \citep{2017MNRAS.466.3060P}.

\cite{2002ApJ...579..800S} and \citet{2008ApJ...685..478N} systematically observed mid-A type stars in the ultraviolet (UV)
looking for the subcoronal emission lines
of C\,{\sc iii} and O\,{\sc vi} formed between 50,000 and 300,000~K \citep{1997A&AS..125..149D,2021ApJ...909...38D}. They find a
very sharp cut-off, where stars with $T_\mathrm{eff}<8200$~K have line fluxes similar to our Sun \citep{1997JGR...102.1641A} when normalized to the bolometric
luminosity, but they claim these lines are undetected in stars with $T_\mathrm{eff} > 8300$~K.
The exact numbers given for $T_\mathrm{eff}$ depend on the effective temperature scale adopted, and will be slightly different in this work.
\citet{2002ApJ...579..800S} conclude that the
transition between stars with and without a corona happens within 100~K and that the cut-off temperature is compatible with theoretical predictions \citep{2000ASPC..210..187C,2002MNRAS.330L...6K}.
\citet{2008ApJ...685..478N} observed O~{\sc vi} in 6 of 8 stars with  $T_\mathrm{eff}>8300$~K, but similar to the X-rays, then present several lines of evidence that the X-ray emission is due to a low-mass companion in all six cases.

\citet{2002ApJ...579..800S} also analyzed archival \emph{ROSAT} observations
and find X-ray emission only for stars that appear below 8200~K on their temperature scale. However, the X-ray sensitivity was not very high by modern standards. 
In this paper, we present new \emph{Chandra}/HRC-I observations that are about an order of magnitude more sensitive for the stars in the \citet{2002ApJ...579..800S} sample that previously only had \emph{ROSAT} data. 
To avoid the problem of unresolved companions, we limit the new observations to well-studied A stars within 30~pc
where spectroscopy and recent planet searches with the radial-velocity method \citep[e.g.][]{2021AJ....161..157H}, direct imaging \citep[e.g.][]{2013ApJ...776....4N,2017AJ....154..245M}, and the study of anomalous GAIA proper motion \citep{2019A&A...623A..72K} essentially rule out the presence of a late-type stellar companion, i.e., our target sample comprises only bona fide single stars.

There are two special classes of A-type stars where X-ray emission is commonly observed: (1) Chemically peculiar Ap \citep{1997A&A...323..121B,2011A&A...531A..58R} stars have magnetic fields presumably strong enough to funnel their stellar wind into the equatorial plane where shocks develop, and (2) pre-main sequence stars \citep[Herbig Ae stars, e.g.][]{2004ApJ...614..221S,2020ApJ...888...15S,2007A&A...468..541T,2009A&A...494.1041G}, some of which even show magnetically collimated jets. Both classes are markedly different in magnetic properties and power sources available for X-ray generation and are not considered any further in this study.

In section~\ref{sec:data}, we show the data from those new observations. We discuss the results in section~\ref{sec:discussion} and summarize our findings in section~\ref{sec:summary}.

\section{Data analysis} \label{sec:data}
We observed four A-types stars with Chandra/HRC-I. Details of the observations are listed in Table~\ref{tab:obslog}. Data was reprocessed with CIAO 4.13 \citep{2006SPIE.6270E..1VF}, following standard analysis procedures.
For reproducibility, we provide the full analysis script\footnote{\url{https://github.com/hamogu/HottestCoolStar/blob/main/figures/HottestCoolStar.ipynb}}.

\begin{table*}
\caption{Chandra observations with pointing information \label{tab:obslog}}
\begin{tabular}{cccccc}
\hline \hline
target & obs. date & OBSID & RA (pointing) & Dec (pointing) & exp. time \\
 &  &  & $\mathrm{{}^{\circ}}$ & $\mathrm{{}^{\circ}}$ & $\mathrm{ks}$ \\
\hline
$\iota$ Cen & 2017-03-31 & 18930 & 200.1525 & -36.7119 & 9.7 \\
$\beta$ Leo & 2017-04-05 & 18931 & 177.2635 & 14.5719 & 10.1 \\
$\delta$ Leo & 2017-02-05 & 18932 & 168.5285 & 20.5240 & 10.1 \\
$\tau^3$ Eri & 2017-06-09 & 18933 & 45.5957 & -23.6223 & 19.9 \\
\hline
\end{tabular}
\end{table*}
First, we try to improve the astrometry of the Chandra observations.
The 90\% uncertainty circle for Chandra absolute astrometry is
0.8\arcsec\footnote{\url{https://cxc.cfa.harvard.edu/cal/ASPECT/celmon/}}. Since
the relative precision is even better than that, the astrometry can be
improved if a sufficient number of sources can be matched to a catalog
with high astrometric precision. We run the CIAO task \texttt{wavdetect} on the
X-ray data to detect X-ray sources with the intent to cross-match them
with 2MASS \citep{2006AJ....131.1163S} or GAIA
\citep{2016A&A...595A...1G,2018A&A...616A...1G}.
In ObsID 18933, we find \object{2MASS J03021318-2335198}, a Seyfert~1 galaxy, to be located 0.7~arcsec to the East of the peak of the X-ray emission; in all other cases we do not find reliable, unambiguous matches or the matched sources are located so far from the aimpoint that their point-spread-function (PSF) is too large to improve coordinate accuracy.

\begin{figure*}
    \centering
    \includegraphics[width=\textwidth]{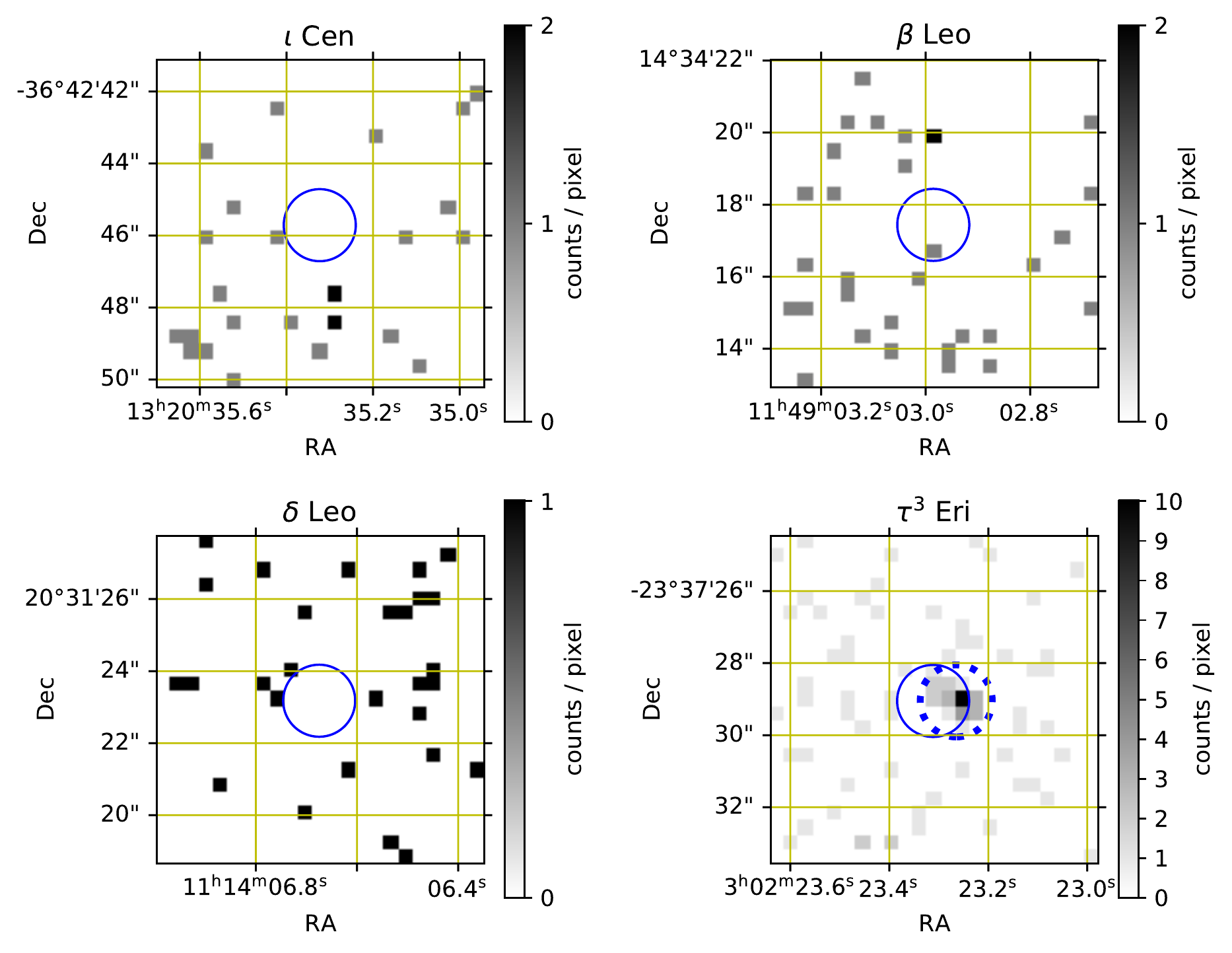}
    \caption{Observations with the Chandra/HRC. The blue circles mark a 1~arcsec radius around the expected position of the target accounting for proper motion at the time of the observation. For $\tau^3$~Eri, the dotted circle accounts for the coordinate offset discussed in the text.
    \label{fig:chandra}}
\end{figure*}

Figure~\ref{fig:chandra} shows the Chandra/HRC-I images of our four
targets. A circle with 1~arcsec radius marks the position of the
target at the time of the observation. Coordinates and proper motions
are taken from \citet{2018yCat.1345....0G} for $\iota$~Cen and from
\cite{2007A&A...474..653V} for the remaining three targets. Only
$\tau^3$~Eri has significant emission within the marked circle in
Fig.~\ref{fig:chandra}. The expected position is about 0.7~arcsec to
the East of the peak of the X-ray emission; the direction and distance of
the offset are very similar to the offset detected in
2MASS~J03021318-2335198 and within the 90\% uncertainty expected for
the Chandra coordinates. We thus conclude that the apparent distance
is in fact due to the uncertainties of the Chandra coordinates and
that the source seen is a detection of $\tau^3$~Eri. 

Next, we determine source flux and uncertainties or upper limits.
We chose a source extraction region with 1.5~arcsec
radius to ensure that the source flux is captured even in the
presence of coordinate uncertainties. The Chandra
PSF depends on the photon energy, but for
soft sources as expected here, that region captures well above 95\%
of the PSF. Without knowing the source spectrum, we cannot fully
correct for the loss of photons outside the source aperture. 
The background flux is determined from
a large region that is apparently source-free; in this way the
statistical error on the background rate is small.

For $\tau^3$~Eri, we calculate 90\% credible intervals on the X-ray flux following a Bayesian approach that
takes into account the presence of a background \citep{2014ApJ...796...24P} as implemented in the CIAO tool \texttt{aprates}. For the three undetected sources, we calculate upper limits following the procedure of \citet{2010ApJ...719..900K}. Deriving an upper limit requires the choice of two parameters. We set the probability of a ``false positive'' (a background fluctuation that is erroneously detected as a source) to be $<0.3$\% corresponding to a
Gaussian-equivalent of ``$3\sigma$'' and the probability of a ``false negative'' (a real source with a true flux above the given upper limit that is not detected because, due to drawing by chance from a Poisson distribution,  the number of photons in the observation is so low that it is compatible with background) to be $<50$\%.

Table~\ref{tab:detections} lists the detected count rate or upper limit. We convert the count rate into an energy flux, assuming a thermal spectrum. For Chandra/HRC-I,
1~ct~ks$^{-1}$ corresponds to an X-ray flux about $1.0\times10^{-14}$~erg~s$^{-1}$ in the 0.1-5~keV band according to WebPIMMS\footnote{\url{https://heasarc.gsfc.nasa.gov/cgi-bin/Tools/w3pimms/w3pimms.pl}}; the variation around this value is no more than 15\% in the temperature range 0.8-13~MK. Using the distance and bolometric luminosity from \citet{2002ApJ...579..800S}, Table~\ref{tab:detections} also lists $L_X$ and $\log(L_X/L_\mathrm{bol})$.
\begin{table*}
\caption{X-ray flux (90\% credible interval) or upper limit (0.3\% false positive and 50\% false negative).\label{tab:detections}}
\begin{tabular}{ccccccc}
  \hline \hline
target & net. counts & net. rate & net. flux & $L_X$ & $\log(L_X/L_\mathrm{bol})$\\
 & $\mathrm{ct}$ & $\mathrm{ct\,ks^{-1}}$ & $10^{-15}\mathrm{erg\,s^{-1}\,cm^{-2}}$ & $10^{26}\mathrm{erg\,s^{-1}}$ \\
\hline
$\iota$ Cen & 0.0 ..  2.9 & 0.0 ..  0.3 & 0 ..   3 &  $<3.2$ & $<-8.4$ \\
$\beta$ Leo & 0.0 ..  3.0 & 0.0 ..  0.3 & 0 .. 2.9 &  $<1.0$ & $<-8.7$ \\
$\delta$ Leo & 0.0 ..  4.8 & 0.0 ..  0.5 & 0 .. 4.7 &  $<2.6$ & $<-8.5$ \\
$\tau^3$ Eri & 20.5 .. 39.8 & 1.0 ..  2.0 & 10 ..   20 & 6 .. 20 & -7.9 .. -7.4 \\
\hline
\end{tabular}
\end{table*}

All target stars are optically bright. We compare their $V$ magnitude
to Vega and scale the number of observed UV events from observations
of Vega \citep{2006ApJ...636..426P}. Based on this we expect 1 or fewer UV
events for each observation and we conclude that UV contamination is
negligible. For $\tau^3$ Eri, we checked the lightcurve and we do not find any significant variability, but --given the low-count number-- even a flare that doubles the X-ray output for a few ks could be hidden in the Poisson noise.

\section{Results and discussion}  \label{sec:discussion}
Our Chandra observations followed up on four targets where \cite{2002ApJ...579..800S} could only give upper limits from \emph{ROSAT} data. We detected $\tau^3$~Eri and pushed the upper limit on $L_X$ for $\delta$~Leo, $\beta$~Leo, and $\iota$~Cen down by about an order of magnitude (Table~\ref{tab:detections}). With that, there is now a one-to-one correspondence between X-ray emission and the transition region lines C\,{\sc iii} and O\,{\sc vi} which are formed between 50,000 and 300,000 K from the sample of \cite{2002ApJ...579..800S}. 
X-ray and C~{\sc iii} 977~\AA{} fluxes are listed in Table~\ref{tab}, which also includes a comparison fluxes and upper limits to other main-sequence stars from the literature.
In the following, discuss implications in the light of other observational or theoretical work that has become available since the \cite{2002ApJ...579..800S} data were published.

\begin{table*}
\caption{Stars of spectral type A with detailed X-ray observations \label{tab}}
\begin{tabular}{ccccccccccc}
\hline\hline
name & age & age ref & $v \sin i$ & $T_\mathrm{eff}$ & $\sigma_{T_\mathrm{eff}}$ & C \sc{iii} 977 \AA{} & $L_\mathrm{X}$ & $\log(L_\mathrm{X}/L_\mathrm{bol})$ & X-ray \\
 & $\mathrm{Myr}$ & ref & $\mathrm{km\,s^{-1}}$ &  K & K & $10^{-7}L_\odot{}$ & $\mathrm{erg\,s^{-1}}$ &  & ref \\
 \hline
HR 4796A & 5-16 & 1,2,3 & 152 &  9750 & 113 &  &  $<1.3\times 10^{27}$ & $< -7.7$ & 16 \\
Vega & 100-500 & 4,5,6 & $25\pm 2$ & 9372 & 503 &  &  $<3.0\times 10^{25}$ & $< -10.0$ & 17\\
$\iota$ Cen & 100-400 & 7,8 & 75 &  9147 & 118 & $<0.10$ & $<3.2\times 10^{26}$ & $< -8.4$ &  here\\
$\beta$ Leo & 30-70 & 9,10,11 & 128 &  8549 & 88 & $<0.03$ & $<1\times 10^{26}$ & $< -8.7$ &  here\\
$\beta$ Pic & 12-40 & 12,13 & 130 &  8103 & 90 &  &  $(1.3\pm0.3)\times10^{27}$ & $-8.2\pm0.1$ & 18 \\
$\delta$ Leo & 600-890 & 7,8 & 180 &  8076 & 155 & $<0.04$ & $<2.6\times 10^{26}$ & $< -8.5$ & here \\
$\tau^3$ Eri & 430-950 & 8 & 133 &  7999 & 103 & $0.96\pm0.16$ & $12_{-6}^{+8}\times10^{26}$ & $-7.6^{+0.2}_{-0.3}$ &  here \\
Altair & 700-1000 & 7,14 & 217 &  7651 & 204 &  & $(1.4\pm0.2)\times10^{27}$ & $-7.4\pm 0.1$ & 19 \\
Alderamin & 1000 & 15 & 196 &  7438 & 173 & $1.34\pm0.13$ & $(2.2\pm0.4)\times10^{27}$ & $-7.5\pm 0.1$ & 20\\
HR 8799 & 38-48 & 9 & 49 &  7187 & 24 &  & $(1.3\pm0.2)\times10^{28}$ & $-6.2 \pm 0.1$ & 21 \\
\hline
\end{tabular}\\
References for column ``age (ref)'':
(1)~\citet{1999ApJ...512L..63W},
(2)~\citet{2013ApJ...767...96W},
(3)~\citet{2014ApJ...786..136D},
(4)~\citet{1998A&A...339..831B},
(5)~\citet{2010ApJ...712..250H},
(6)~\citet{2010ApJ...708...71Y},
(7)~\citet{2012AJ....143..135V},
(8)~\citet{2015ApJ...804..146D},
(9)~\citet{2015MNRAS.454..593B},
(10)~\citet{2019ApJ...870...27Z},
(11)~\citet{2019MNRAS.489.2189L},
(12)~\citet{2001ApJ...562L..87Z},
(13)~\citet{2010ApJ...723.1599M},
(14)~\citet{2018AJ....156..286S},
(15)~\citet{2009ApJ...701..209Z}. 
Values for $v \sin i$ are taken from the compilation of \citet{2002A&A...393..897R} (see there for references for individual targets). Only for Vega is the value for $v \sin i$ measured in that work and an uncertainty provided. $\iota$~Cen is not part of that catalog, and so the value given in \citet{2002ApJ...579..800S} is used instead. $T_\mathrm{eff}$ is quoted from the PASTEL catalog (\citealt{2016A&A...591A.118S}; see there for references for individual targets), but see Section~\ref{sec:discussion} for discussion on why  $T_\mathrm{eff}$ may not characterize the photospheric properties well. Fluxes and upper limits for C~{\sc iii} 977 \AA{} are taken from \citet{2002ApJ...579..800S}.
References for column ``X-ray (ref)'':
(16)~\citet{2014ApJ...786..136D},
(17)~\citet{2006ApJ...636..426P}
(18)~\citet{2012ApJ...750...78G},
(19)~\citet{2009A&A...497..511R},
(20)~\citet{2002ApJ...579..800S},
(21)~\citet{2010A&A...516A..38R}.
\end{table*}

Note that ages given in Table \ref{tab} are often based on membership in moving groups. In some cases, membership is under debate and stars could be considerably older if they turn out to be field stars, e.g., see the discussion and references in \citet{2021AJ....161..186D} for $\beta$~Leo. Table~\ref{tab} also gives an effective temperature for the stars.
No single survey covers all stars in Table~\ref{tab} and it is well known that different classification methods lead to discrepant spectral types \citep[e.g.,][]{1989ApJS...70..623G}.
Additionally, the stellar surface of A-type stars is not necessarily well described by a single spectral type or temperature; e.g.\ \citet{2010A&A...516A..38R} argued that the
spectral classification of HR~8799 of A5 is based on metal lines,
while the atmospheric temperature might be better characterized by its
hydrogen lines, which would make it an F0 star
\citep{1999AJ....118.2993G}, cool enough to generate magnetic activity
through a convective dynamo. Similarly,  Altair is a very fast rotator, and thus it has a considerable equatorial bulge and consequently a lower temperature on the equator than on the pole, which might allow the formation of a thin convection zone in the equatorial bulge \citep{2009A&A...497..511R}. 
The $T_\mathrm{eff}$ given in the table is the average of all values listed in the PASTEL catalog (\citealt{2016A&A...591A.118S}; see there for references for individual targets), which collects high precision spectroscopic and photometric measurements of the effective temperature. Where more than one measurement is available in PASTEL, the $\sigma_{T_\mathrm{eff}}$ is calculated as the standard deviation of all measurements, since the systematic differences between different input data are typically larger than individual measurement uncertainties and uncertainties are not available in all cases. We stress that $T_\mathrm{eff}$ only provides an approximate measure of the photospheric temperature in the region where UV and X-rays are generated, and large temperature gradients can exist between pole and equator.
Thus, the ordering in Table~\ref{tab} by $T_\mathrm{eff}$ from hot to cool should be regarded as approximate.

Numbers for the rotational velocity $v \sin i$ are taken from a compilation \citep{2002A&A...393..897R}, except for $\iota$ Cen which is not part of the list, so we use the value from \cite{2002ApJ...579..800S}.

\subsection{Comparing X-ray and UV fluxes}
\begin{figure}
    \centering
    \includegraphics[width=.4\textwidth]{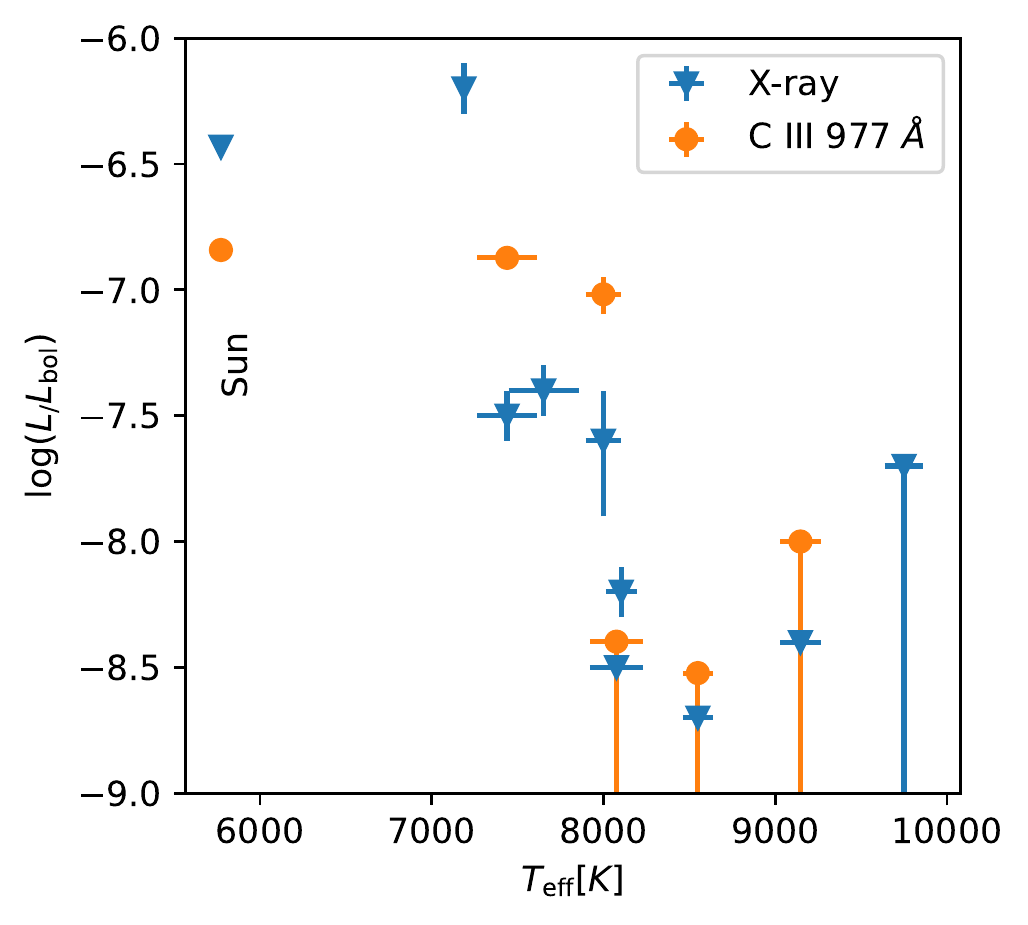}
    \caption{X-ray and C~{\sc iii} 977~\AA{} line flux. Both fluxes are shown relative to the star's bolometric luminosity. Stars cooler than about 8000~K are all detected in X-rays and  C~{\sc iii} 977~\AA{}, while stars that are significantly hotter are at least an order of magnitude fainter.
The solar data is taken from \citep{1997JGR...102.1641A}, the other UV fluxes from \citet{2002ApJ...579..800S}. } \label{fig:lc3lxlbol}
\end{figure}

Table~\ref{tab} lists one UV line, C~{\sc iii} 977~\AA{}, but \citet{2002ApJ...579..800S} show that all stars where that line can be detected also show detections in C~{\sc iii} 1175~\AA{} and the O~{\sc vi} 1032/1038~\AA{} doublet; similarly, all stars where there is only an upper limit on C~{\sc iii} 977~\AA{} also only have an upper limit on the other lines. 
Figure~\ref{fig:lc3lxlbol} compares the X-ray and the C~{\sc iii} 977~\AA{} line fluxes as a fraction of the bolometric flux. The three X-ray undetected stars ($\iota$~Cen, $\delta$~Leo, and $\beta$~Leo) are at least an order of magnitude fainter than $\tau^3$~Eri and Alderamin, which in turn are already significantly fainter than the Sun. So, all studied A-type stars show comparatively less chromospheric and coronal activity. 
If this was an effect of the coronal temperature, where A-type stars have cooler coronae than the Sun, then we would expect additional emission in the UV, yet this does not seem to be the case as the C~{\sc iii} and O~{\sc vi} (not shown in the figure) also drop as $T_\mathrm{eff}$ increases. The measurements instead indicate that the relative amount of emitting plasma decreases. This could mean that a lower fraction of the surface area is covered with active structures, or that those structures have lower densities or shorter lifetimes than in later-type stars.

At first sight, one might expect also a correlation with rotational velocities, where faster rotation produces a stronger X-ray activity as is clearly seen in lower mass main-sequence stars \citep{2011ApJ...743...48W}, but Table~\ref{tab} does not bear that out. $\tau^3$~Eri is seen in X-rays, while $\delta$~Leo and $\beta$~Leo are equally fast rotators, but slightly hotter and not detected in X-rays. 
For most stars in Table~\ref{tab} the rotational period is measured or can be inferred assuming that the stellar spin axis is aligned with the axis of a debris disk or planetary system. The stars not detected in X-rays are fast rotators, e.g.\ Vega \citep[0.68~d,][]{2015A&A...577A..64B} and HR~4796A,  \citep[0.5~d,][]{2014ApJ...786..136D}, as are the stars that are detected in X-rays, e.g.\ Alderamin \citep[0.5~d,][]{2006ApJ...637..494V} and Altair \citep[0.4~d,][]{2006ApJ...636.1087P}. Thus,
the deciding factor for detectability of X-ray and transition region UV is not the rotation period, but seems to be the spectral type or photospheric temperature.

\subsection{Comparison with earlier and later stars}
\begin{figure*}
    \centering
    \includegraphics[width=\textwidth]{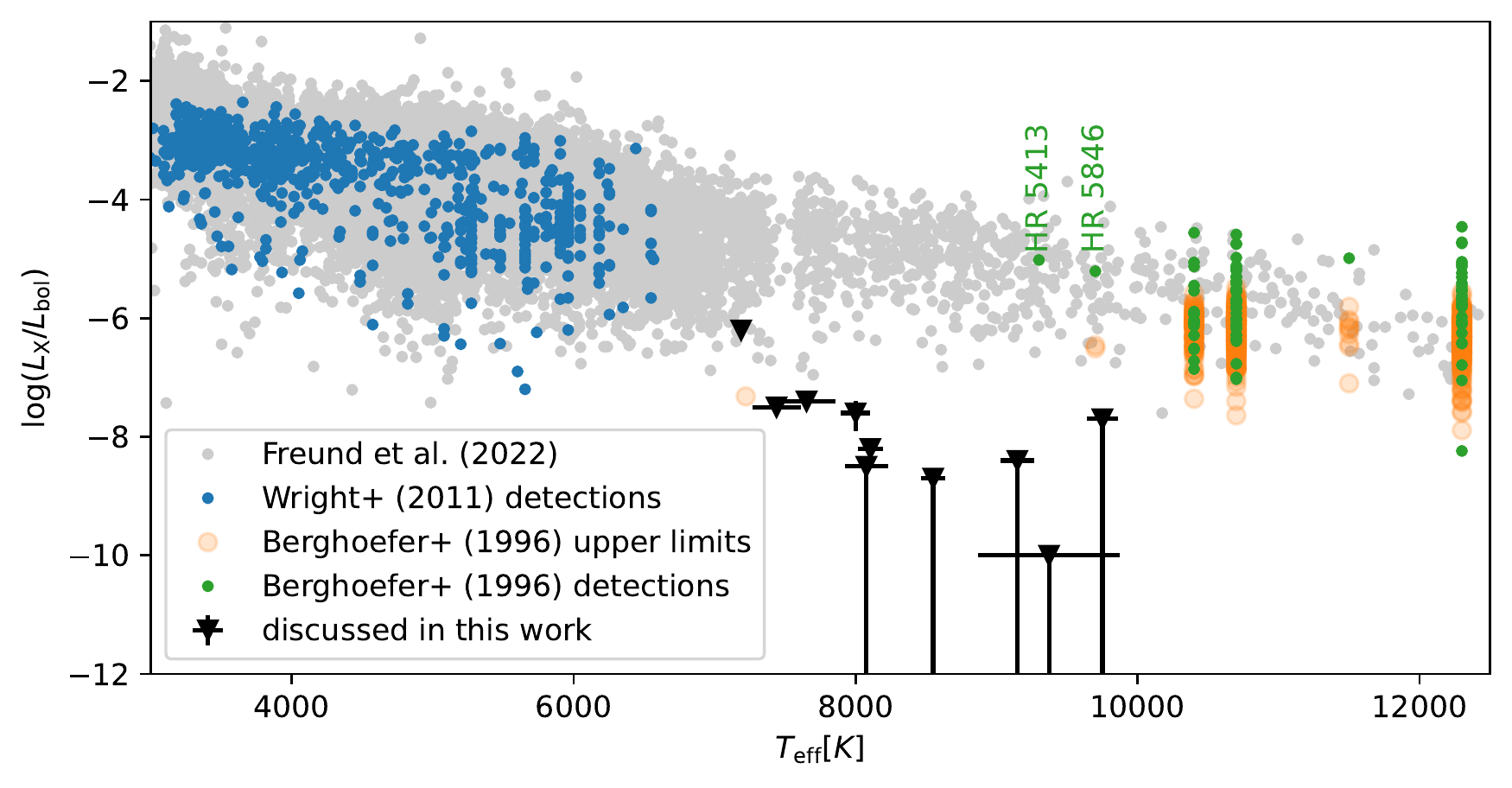}
    \caption{Comparison of the X-ray luminosity in A-type stars with data for earlier and later spectral types. Early A-type stars are fainter in X-rays than other spectral types.
     Note that the ROSAT catalog compiled by \citet{Freund} does not resolve most binaries and thus especially the data points listed in the range $T_\mathrm{eff}=7500-10000$~K represent the $L_\mathrm{bol}$ from an A-star primary and the $L_X$ from an unresolved late-type companion. This highlights the importance of carefully selecting a sample of single A stars as done in this work (black points).} \label{fig:lxlbol}
\end{figure*}

Figure~\ref{fig:lxlbol} compares A stars from Table~\ref{tab} with earlier and later-type stars. We use the effective temperature as common axis for all datasets. 
O and B stars are taken from \citet{1996A&AS..118..481B} who distinguish between detections and upper limits. As it turns out, the upper limits are mostly in the same range of $L_X/L_\mathrm{bol}$ as the detected sources; they might just be undetected because they are further away or observed for a shorter time. To convert the spectral types of all those stars into an effective temperature, we use the compilation in \citet{2013ApJS..208....9P}. While that list is valid only for dwarf stars and a number of stars here have other luminosity classes, the conversion between color and temperature is sufficient for the purpose of this plot.

The X-ray activity in cool stars was studied by \citet{2011ApJ...743...48W}. In cool stars, magnetic activity scales with Rossby number or rotation rate, so there is considerable scatter if shown as a function of effective temperature. Very few stars in the sample are listed with spectral types, so instead we use the V-K color and convert it to a temperature, again using the table from \citet{2013ApJS..208....9P}. 
We also compare the new A star data with the stellar sources from the ROSAT survey identified by \citet{Freund} who matched ROSAT and Gaia sources using a Bayesian method to select bona fide stellar X-ray sources. While their formal sample reliability is above 90\% (i.e.\ at least 9 out of 10 sources are correctly identified as stellar), almost all companions remain unresolved in X-rays. Therefore, in many cases the ROSAT detected X-ray emission is not from the primary seen by Gaia, especially for A stars which are very often found in binary/multiple systems. This highlights that for detailed studies of A-star stellar activity, as we do in this work, it is important to select the sample carefully.

The fractional X-ray flux drops with increasing temperature for late-type stars. The brightest of the A stars (HR 8799) might follow that same trend, but the others are considerably less luminous in X-rays, indicating a qualitative change in the mechanism that generates the magnetic field and powers the X-ray emission. In particular, the early A stars (most notably Vega) have upper limits on the X-ray flux orders of magnitude below the later type stars. 

Figure~\ref{fig:lxlbol} also shows two stars with effective temperature from \citet{1996A&AS..118..481B} very similar to our hottest stars (Vega and HR~4796A) in effective temperature, but as bright in X-rays as typical later type stars or the brightest B-type stars. However, HR~5413 (HIP 70931) has a low-mass companion ($0.23\;M_\sun$) at 0.6~arcsec resolved with adaptive optics imaging \citep{2014MNRAS.437.1216D} which might well be the X-ray source and for HR~5846 (HIP 77086) a low-mass companion is indicated by a Gaia proper-motion anomaly \citep{2019A&A...623A..72K}. The mass estimate for the companion depends on the orbital separation, but this might well be an X-ray active low-mass star. These two examples demonstrate why a study of A-star X-ray activity requires a sample of well studied targets with strict limits on binarity.


\subsection{Coronal and chromospheric activity}
\cite{2002ApJ...579..800S} detect the UV emission lines
of C~{\sc iii} at 977\AA{} and 1175\AA{} as well as the O~{\sc vi}
doublet 1032/1037\AA{} in single stars up to spectral type A4 (which
they associate with $T_\mathrm{eff}=8200$~K), but not in hotter stars. They also see those
lines in $\beta$~Ari, which has a primary component hotter than this
limit, but conclude that the most likely origin for the observed
emission is chromospheric activity in the cooler secondary in the
system. The emission lines of C~{\sc iii} and O~{\sc vi} are formed
below the corona; they have peak formation temperatures between
50,000~K and 300,000~K. Combining our new, more sensitive Chandra data with the X-ray data already discussed in \cite{2002ApJ...579..800S}, 
we now see a one-to-one correspondence between X-ray emission and
C~{\sc iii} and O~{\sc vi} lines in the single stars. This confirms that corona and chromosphere are powered by the same mechanism.
 As stars become hotter, the convective envelope shrinks and at some point the dynamo mechanism breaks down.
 
 Within the uncertainties $\tau^3$~Eri, $\beta$~Pic, and
$\beta$~Leo could have the same $T_\mathrm{eff}$, yet the latter is undetected in X-rays with an upper limit a few times below the $L_X/L_{bol}$ and  $L_\mathrm{C {\sc III}}/L_{bol}$ (Figure~\ref{fig:lc3lxlbol}) for $\tau^3$~Eri and $\beta$~Pic. This indicates that
the drop in X-ray and UV activity is very sharp and occurs within $\Delta T_\mathrm{eff}<200$~K. The exact boundary could depend on other factors, such as age or metallicity, which would have to be probed by a larger sample.

\subsection{A-star  X-ray emission: Debris disks, planets, or spectral type?}
Debris disks have been found in 
HR~4796A \citep[e.g.][]{1991ApJ...383L..79J}, Vega
\citep[e.g.][]{2005ApJ...628..487S}, $\iota$~Cen
\citep[e.g.][]{2011ApJ...736L..32Q}, $\beta$~Leo  \citep[e.g.][]{2021AJ....161..186D}, and
$\beta$~Pic \citep[e.g.][]{2001MNRAS.323..402L}.  These debris disks
form when planets or planetesimals in orbit around the star
collide. The smaller dust grains are blown out of the system through
radiation pressure relatively quickly and need to be replenished
constantly by grinding down larger objects. 
Some stars with debris disks have X-ray and transition region UV emission, others do not, so we conclude that the presence of debris disks is not the deciding factor for chromospheric or coronal activity.

Since they are close and bright targets, the stars in Table~\ref{tab}
have also been targeted by planet searches. Massive planets are
confirmed for HR~8799 \citep{2008Sci...322.1348M} and $\beta$~Pic
\citep[e.g.][]{2021AJ....161..179B}, but they are located at several
AU distance from the star and are thus unlikely to influence the
X-ray emission from the star.

For late-type stars with a convective dynamo, X-ray activity can be parameterized by the Rossby number, or the ratio of the rotation period $P_{\rm rot}$ to the convective turnover time $\tau_{\rm conv}$, i.e.
\begin{equation}
{\rm Ro} \sim \frac{P_{\rm rot}}{\tau_{\rm conv}}. \label{eqn:ro}
\end{equation}
A higher Rossby number is associated with lower X-ray activity across pre-main sequence, main sequence, and giant stars \citep{Preibisch+2005, Pizzolato+2003, Gondoin+2005,2011ApJ...743...48W} and established in stellar dynamo models \citep[e.g.][]{Brandenburg+1998}.

Given that many stars in Table~\ref{tab} have rotation periods less than a day, the decline of the X-ray emission, and thus (in the scheme of a solar-like dynamo) increasing Rossby number, then points to a rapidly declining $\tau_{\rm conv}$ with increasing photospheric temperature.

\subsection{How could a dynamo operate in A-type stars?}
\label{sect:dynamos}

X-ray activity is common in later-type stars and for Altair
\citet{2009A&A...497..511R} concluded that the X-ray activity is
probably concentrated in the equatorial region, which is also
responsible for the generation of the chromosphere
\citep{1995ApJ...439.1011F}. Similar to Altair, $\tau^3$~Eri is also a
rapid rotator with $v\sin i=180$~km~s$^{-1}$ and an estimated oblateness
(depending on the inclination angle $i$) of around 7\%
\citep{2012A&ARv..20...51V}. On the other hand, $\delta$~Leo is an
equally fast rotator and remains undetected.

\citet{2019ApJ...883..106C} suggest the presence of 
convection zones in A and late B stars very close to the surface. These thin zones are caused
by partial ionization of H and He. For A-type stars about $10^{-2}-10^{-3}$ of the total $L_\mathrm{bol}$ is transported through convective motion. Assuming equipartition between kinetic and magnetic energy, only $10^{-4}$ of the available magnetic energy would have to be converted into X-ray flux to power the observed X-ray luminosity. However, since the structure of those fields would differ markedly from the Sun, coronal heating would also look very different. Furthermore, \citet{2019ApJ...883..106C} predict that this mechanism should operate to mid-B type stars, in contrast to the drop of $L_X$/$L_{bol}$ that we observe in mid-A type stars in our sample.

\citet{2014ApJ...786..136D} extensively discuss two types of shear
dynamos that could transform some initial differential rotation of a
young star into a magnetic field that quickly decays, on time scales
of order of a Myr for a dynamo based on magnetic buoyancy
\citep{1995MNRAS.272..528T} or on times scales of order 300~yr for a
dynamo based on the Taylor instability
\citep{2002A&A...381..923S,2006A&A...449..451B}. Both scenarios
involve physical parameters that are uncertain by orders of magnitude
and \citet{2014ApJ...786..136D} use their upper limit on the X-ray
flux from HR~4796A to exclude the most optimistic values for those
parameters. Our limit for $\beta$~Leo is significantly stricter, but
the star is also older (Table~\ref{tab}), which places about the same
limits on the parameters for the \citet{1995MNRAS.272..528T} model as
does HR~4796A \citep[See Fig 3 in][]{2014ApJ...786..136D}.

Rapid rotation leads to a bulged and cooler equator in these stars \citep[on Altair][resolved a temperature difference between pole and equator of at least 1000~K]{2007Sci...317..342M}, so one may speculate that some of the stars might be conducive to a convective instability near their equators as has been suggested by \citet{2009A&A...497..511R} for Altair. If so, then a convective dynamo near the equator may sustain significant levels of magnetic activity. Since the X-ray activity is becoming appreciable starting at $\tau^3$~Eri, it suggests that $\tau^3$~Eri and cooler stars are in an appropriate regime to host equatorial convection. 

While the X-ray emitting corona is magnetically heated, the chromosphere in late-type stars has both acoustically and magnetically heated components \citep{1999ApJ...522.1053C,2000ssma.book.....S}, where the outermost and hottest layers of the chromosphere require magnetic heating \citep{2002A&A...386..983F}. The UV lines are formed at the high end of the temperature range of the chromosphere, and thus they likely depend on magnetic heating like the corona. Our observational result that UV lines and X-rays are visible in the same objects confirms this idea.

\section{Summary}
\label{sec:summary}
We present Chandra/HRC-I observations of four early A stars. $\tau^3$~Eri is clearly detected, while we set sensitive upper limits on the other three targets. With this detection and our new upper limits that are an order of magnitude better than previous \emph{ROSAT} data, there is now a one-to-one correspondence between X-ray emission and C\,{\sc iii} and O\,{\sc vi} lines, which are formed between 50,000 and 300,000 K in the transition region and were observed by \citet{2002ApJ...579..800S}. This confirms that both regions are powered by the same physical mechanism and that this mechanism essentially switches off around $T_\mathrm{eff}=8100\pm200$~K as we observe a drop of X-ray luminosity by at least an order of magnitude. Given the rather old age of $\tau^3$~Eri, which we detect in X-rays, the magnetic field that powers X-ray emission and UV lines needs to be generated continuously. We discuss different dynamo mechanisms that could power the observed X-ray emission, but we are unable to conclusively select any particular one.

\begin{acknowledgements}

This research has made use of data obtained from the Chandra Data Archive, and software provided by the Chandra X-ray Center (CXC) in the application package CIAO.
This research has made use of the SIMBAD database,
operated at CDS, Strasbourg, France \citep{2000A&AS..143....9W}. 
This research has made use of the VizieR catalogue access tool, CDS,
 Strasbourg, France (DOI : 10.26093/cds/vizier). The original description 
 of the VizieR service was published in \citet{2000A&AS..143...23O}.
This research has made use of NASA’s Astrophysics Data System Bibliographic Services.
This publication makes use of data products from the Two Micron All Sky Survey, which is a joint project of the University of Massachusetts and the Infrared Processing and Analysis Center/California Institute of Technology, funded by the National Aeronautics and Space Administration and the National Science Foundation.
This work has made use of data from the European Space Agency (ESA) mission
{\it Gaia} (\url{https://www.cosmos.esa.int/gaia}), processed by the {\it Gaia}
Data Processing and Analysis Consortium (DPAC,
\url{https://www.cosmos.esa.int/web/gaia/dpac/consortium}). Funding for the DPAC
has been provided by national institutions, in particular the institutions
participating in the {\it Gaia} Multilateral Agreement.
HMG was supported by the National Aeronautics and Space Administration through Chandra Award Number GO9-20018X issued by the Chandra X-ray Observatory Center, which is operated by the Smithsonian Astrophysical Observatory for and on behalf of the National Aeronautics Space Administration under contract NAS8-03060. C.M.\ acknowledges support from NASA ADAP grant 18-ADAP18-0233. S.J.W.\ was supported by the Chandra X-ray Observatory Center, which is operated by the Smithsonian Astrophysical Observatory for and on behalf of the National Aeronautics Space Administration under contract NAS8-03060. J.R. acknowledges support from the DLR under grant 50QR2105.

\end{acknowledgements}

\facilities{Chandra/HRC}

\software{AstroPy \citep{2013A&A...558A..33A,2018AJ....156..123A},
astroquery \citep{2019AJ....157...98G},
CIAO \citep{2006SPIE.6270E..1VF}, NumPy \citep{van2011numpy,harris2020array}, Matplotlib \citep{Hunter:2007}, Sherpa \citep{2007ASPC..376..543D,doug_burke_2021_4428938}}

\bibliography{bib}{}
\bibliographystyle{aasjournal}



\end{document}